\documentclass[reqno,11pt]{amsart}

\usepackage{epsf,amssymb}

\def\be{\begin{equation}}
\def\ee{\end{equation}}
\def\ba{\begin{align}}
\def\bm{\begin{multline}}
\def\bfig{\begin{figure}[htb]}
\def\efig{\end{figure}}

\newcommand{\paper}[1]{{\it #1}, }
\newcommand{\journal}[4]{#1 #2, #3 (#4)}
\newcommand{\CMP}{Commun.\ Math.\ Phys.}
\newcommand{\HPA}{Helv.\ Phys.\ Acta}
\newcommand{\JSP}{J.\ Stat.\ Phys.}
\newcommand{\RMP}{Rev.\ Math.\ Phys.}

\numberwithin{equation}{section}
\newtheorem{theorem}{Theorem}[section]
\newtheorem{proposition}[theorem]{Proposition}
\newtheorem{lemma}[theorem]{Lemma}

\newcommand{\nn}{\nonumber}

\renewcommand{\leq}{\leqslant}
\renewcommand{\geq}{\geqslant}

\newcommand{\dd}{{\rm d}}
\newcommand{\e}[1]{{\rm e}^{#1}}

\newcommand{\sumtwo}[2]{\sum_{\substack{#1 \\ #2}}}
\newcommand{\sumthree}[3]{\sum_{\substack{#1 \\ #2 \\ #3}}}

\newcommand{\Lowerint}[1]{\lfloor #1 \rfloor}

\def\Tr{{\operatorname{Tr\,}}}

\def\bra #1{\langle#1 |\,}
\def\ket #1{\,|#1 \rangle}

\newcommand{\upchi}{\raise 2pt \hbox{$\chi$}}

\def\writefig#1 #2 #3 {\rlap{\kern #1 truecm \raise #2 truecm
\hbox{#3}}}
\def\figtext#1{\smash{\hbox{#1}} \vspace{-5mm}}

\newcommand{\caA}{{\mathcal A}}

\newcommand{\caC}{{\mathcal C}}

\newcommand{\caF}{{\mathcal F}}
\newcommand{\caG}{{\mathcal G}}
\newcommand{\caH}{{\mathcal H}}

\newcommand{\caL}{{\mathcal L}}

\newcommand{\bbA}{{\mathbb A}}

\newcommand{\bbN}{{\mathbb N}}

\newcommand{\bbR}{{\mathbb R}}

\newcommand{\bbZ}{{\mathbb Z}}

\newcommand{\bsn}{{\boldsymbol n}}

\begin{document}


\title{Mott transition in lattice boson models}

\author{R.\ Fern\'andez, J.\ Fr\"ohlich, D.\ Ueltschi}

\address{Roberto Fern\'andez \hfill\newline
Laboratoire de Math\'ematiques Rapha\"el Salem \hfill\newline
UMR 6085 CNRS-Universit\'e de Rouen \hfill\newline
Avenue de l'Universit\'e, BP.12 \hfill\newline
F--76821 Saint Etienne du Rouvray, France \newline\indent
{\rm http://www.univ-rouen.fr/LMRS/Persopage/Fernandez/}}
\email{Roberto.Fernandez@univ-rouen.fr}

\address{J\"urg Fr\"ohlich \hfill\newline
Institut f\"ur Theoretische Physik \hfill\newline
Eidgen\"ossische Technische Hochschule \hfill\newline
CH--8093 Z\"urich, Switzerland \newline\indent
{\rm http://www.itp.phys.ethz.ch/mathphys/juerg}}
\email{juerg@itp.phys.ethz.ch}

\address{Daniel Ueltschi \hfill\newline
Department of Mathematics \hfill\newline
University of Arizona \hfill\newline
Tucson, AZ 85721, USA \newline\indent
{\rm http://math.arizona.edu/$\sim$ueltschi}}
\email{ueltschi@math.arizona.edu}

\renewcommand{\thefootnote}{}
\footnote{$^*$Collaboration supported in part by the Swiss National Science Foundation
under grant 2-77344-03.}
\setcounter{footnote}{0}
\renewcommand{\thefootnote}{\arabic{footnote}}

\maketitle

\begin{quote}
{\small

We use mathematically rigorous perturbation theory to study the transition between the Mott insulator
and the conjectured Bose-Einstein condensate in a hard-core Bose-Hubbard model. The critical line is established to lowest order in the tunneling amplitude.

\vspace{1mm}

}  

\vspace{1mm}

\vspace{1mm}

\noindent
{\footnotesize\it 2000 Math.\ Subj.\ Class.: 82B10, 82B20, 82B26}

\end{quote}

\section{Introduction}
\label{secintro}

Initially introduced in 1989 \cite{FWGF}, the Bose-Hubbard model has been the object of
much recent work. It represents a simple lattice model of itinerant
bosons which interact locally. This model turns out to describe fairly well recent
experiments with bosonic atoms in optical lattices \cite{GMEHB,KMSSE}. Its low-temperature
phase diagram has been uncovered in
several studies, both analytical (see e.g.\ \cite{FWGF,FM, EM}) and numerical
\cite{BASD,STTD} ones.
When parameters such as the chemical potential or the tunneling amplitude are varied the
Bose-Hubbard model exhibits a phase transition from a Mott insulating phase to a Bose-Einstein
condensate. Fig.\ \ref{figphdBH}, below, depicts its ground state phase diagram.

In this paper, we investigate the phase diagram of this model in a mathematically rigorous way. We focus
on the situation with a small tunneling amplitude, $t$, and a small chemical potential,
$\mu$. We construct the critical line between Mott and non-Mott behavior to lowest
order in the ratio $t/\mu$. More precisely, we prove the existence of domains with and without Mott insulator. These domains are separated by a comparatively thin stretch; the domain without Mott insulator is widely believed to be a Bose condensate. Our results establish in particular the occurrence of a ``quantum phase transition'' in the ground state.

Over the years several analytical methods have been developed that are useful for the
study of models such as the Bose-Hubbard model. They include a general theory of classical
lattice systems with
quantum perturbations \cite{BKU,DFF,DFFR,KU}. These methods can be used to establish the existence of Mott phases for
small $t$; but they only apply to domains of parameters far from the transition lines.
The Bose-Hubbard model on the complete graph can be studied rather explicitly and its phase diagram is
similar to the one of the finite-dimensional model \cite{BD}.
Results using reflection positivity are mentioned below and only apply to the hard-core
model. A related model with an extra chessboard potential was studied in \cite{ALSSY} (see
also \cite{LSSY}).

The Bose-Hubbard model is defined as follows. Let $\Lambda \subset \bbZ^d$ be a finite
cube of volume $|\Lambda|$. We introduce the bosonic Fock space
\be
\caF_\Lambda = \bigoplus_{N\geq0} \caH_{\Lambda,N},
\ee
where $\caH_{\Lambda,N}$ is the Hilbert space of symmetric complex functions on $\Lambda^N$.
Creation and annihilation operators for a boson at site $x\in\Lambda$ are denoted by
$c^\dagger_x$ and $c_x$, respectively. The Hamiltonian of the Bose-Hubbard model is given
by
\be
\label{defHam}
H = -t \sumtwo{x,y\in\Lambda}{|x-y|=1} c^\dagger_x c_y + \tfrac12 U \sum_{x\in\Lambda}
c^\dagger_x c_x \bigl( c^\dagger_x c_x - 1 \bigr)
\ee
The first term in the Hamiltonian represents the kinetic energy; the
hopping parameter $t$ is chosen to be positive. The second term is an on-site
interaction potential (assuming each particle interacts with all other particles at the
same site). The interaction is proportional to the number of pairs of particles; the
interaction parameter $U$ is positive, and this corresponds to repulsive
interactions. In our construction of the equilibrium state, we work in the grand-canonical
ensemble. This amounts to adding a term $-\mu
N$ to the Hamiltonian, where $N = \sum_x c^\dagger_x c_x$ is the number operator, and $\mu$ is the chemical
potential.

The limit $U\to\infty$ describes the {\it hard-core Bose gas} where each site can be
occupied by at most one particle. This model is equivalent to the $xy$ model with spin
$\frac12$ in a magnetic field proportional to $\mu$.
Spontaneous magnetization in the spin model corresponds to Bose-Einstein condensation
in the boson model. The presence of a Bose condensate has been rigorously established for
$\mu=0$ (the line of hole-particle symmetry). See \cite{DLS} for a proof valid at low
temperatures in three dimensions, and \cite{KLS} for an analysis of the ground state in two dimensions.
The proofs exploit reflection positivity and infrared bounds, a method that was originally introduced for the classical
Heisenberg model in \cite{FSS}. At present, there are no
rigorous results about the presence of a condensate for $\mu\neq0$, or for finite $U$.

\bfig
\epsfxsize=100mm
\centerline{\epsffile{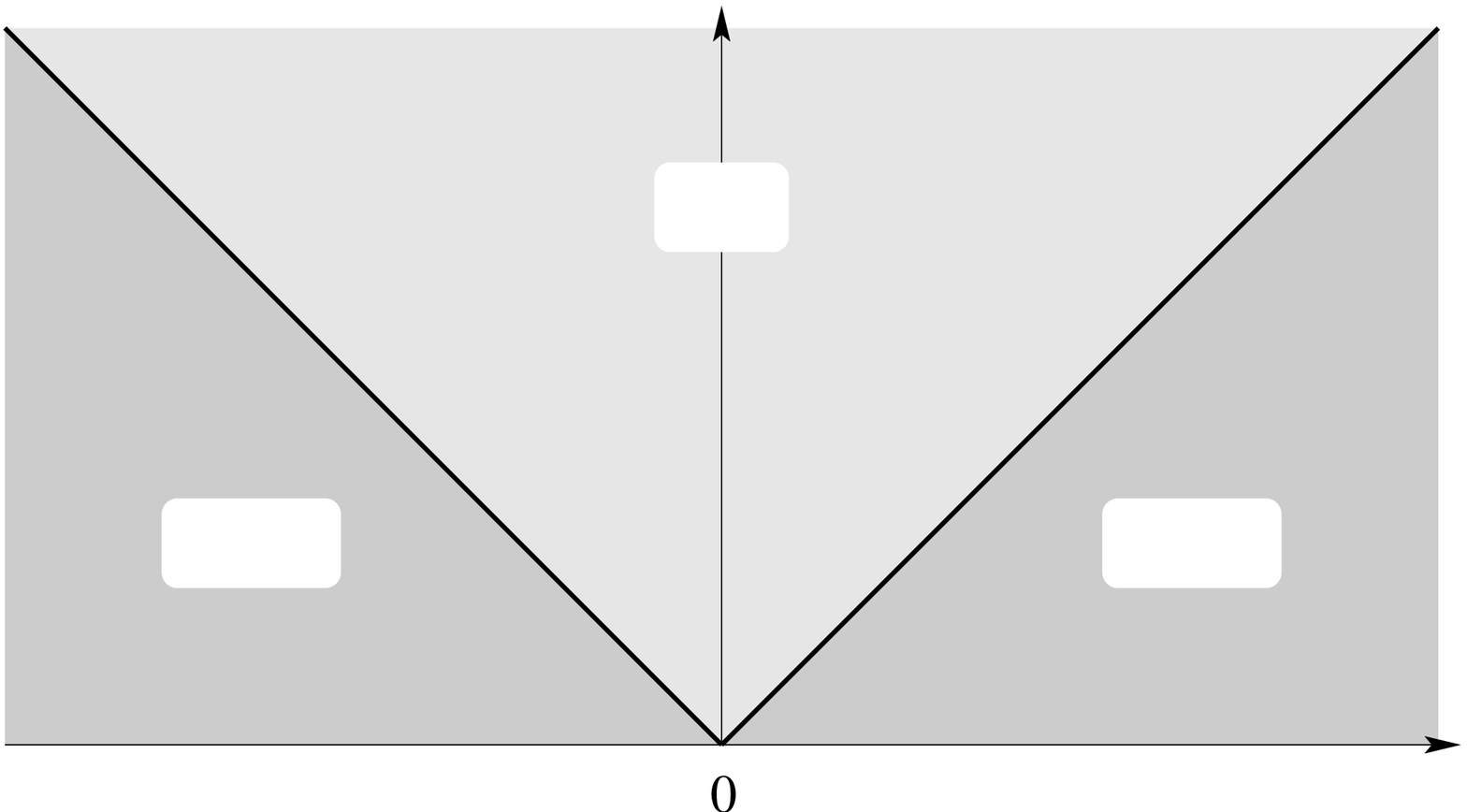}}
\figtext{
\writefig	11.9	0.7	{$\mu$}
\writefig	6.4	5.7	{$t$}
\writefig	3.6	4.2	{$-\frac\mu{2d}$}
\writefig	9.5	4.2	{$\frac\mu{2d}$}
\writefig	6.4	4.45	{BEC}
\writefig	3.1	2.2	{$\rho=0$}
\writefig	9.55	2.2	{$\rho=1$}
}
\caption{Zero temperature phase diagram for the hard-core Bose gas.
Bose-Einstein condensation is proved on the line $\mu=0$, for any $t>0$. Our perturbation
methods provide a quantitative description of the Mott insulator phases with density 0 and 1.}
\label{figphdhcBH}
\end{figure}

The ground state phase diagram of the hard-core Bose gas is depicted in Fig.\
\ref{figphdhcBH} and reveals three regions: a phase with empty sites, a phase with
Bose-Einstein condensation in dimension greater or equal to two, and a phase with full occupation. Particle-hole symmetry
implies that the phase diagram is symmetric around the axis $\mu=0$.

The critical value of the hopping parameter in the ground state of the hard-core (hc) Bose
gas is
\be
t_{\rm c}^{\rm hc}(\mu) = \frac{|\mu|}{2d}.
\ee
This follows by observing that the cost of adding one particle in a state of vanishing
density (where interactions are negligible) is $-\mu-2dt$. For $\mu<-2dt$ the empty
configuration minimizes the energy, while for $\mu>-2dt$ a state with sufficiently low,
but positive density has
negative energy. The Mott phases of the hard-core Bose gas at zero temperature are stable
because of the
absence of `quantum fluctuations' --- the ground state is just the empty or the full
configuration. The hard-core model is an excellent approximation to the
general Bose-Hubbard model when $t$ is small and $\mu$ is sufficiently small.

A first insight into the ground state phase diagram of the general Bose-Hubbard model is
obtained by restricting the Hamiltonian to low energy configurations. Namely, for
$-\infty < \mu \lesssim \frac12 U$, low energy states have 0 or 1 particle per site. The
restricted model is the hard-core Bose gas. Next, for $\frac12 U \lesssim \mu
\lesssim \frac32 U$, states of lowest energy have 1 or 2 particles per site. The restricted model
is again a hard-core Bose gas, but with effective hopping equal to $2t$. We can define
projections onto subspaces of low energy states for all $\mu$; corresponding restricted
models yield the following approximation for the critical hopping parameter:
\be
\label{tcritapprox}
t_{\rm c}^{\rm approx}(\mu) = \begin{cases} \frac{|\mu|}{2d} & \text{if } -\infty < \mu <
\frac12 U, \\ \frac{|\mu-kU|}{2d(k+1)} & \text{if } (k-\frac12)U < \mu < (k+\frac12)U, \,
k\geq1, \end{cases}
\ee
(thin lines in Fig. \ref{figphdBH}). The true critical line $t_{\rm c}(\mu)$
agrees with $t_{\rm c}^{\rm approx}(\mu)$ up to corrections due to quantum fluctuations.
We expect that
\be
t_{\rm c}(\mu) = t_{\rm c}^{\rm approx}(\mu) \bigl( 1 + O(\tfrac tU) \bigr).
\ee
\bfig
\epsfxsize=125mm
\centerline{\epsffile{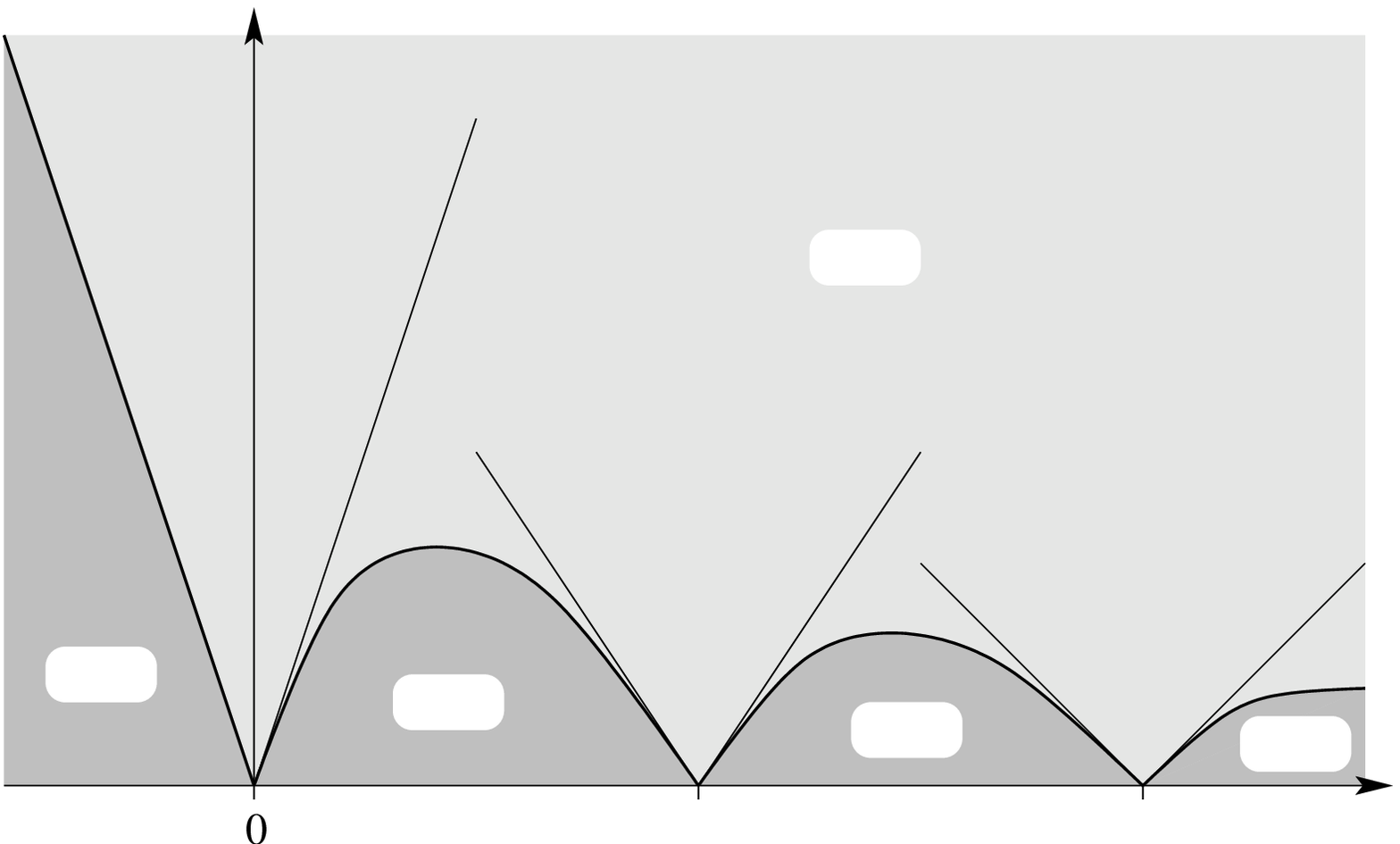}}
\figtext{
\writefig	13.0	0.7	{$\mu$}
\writefig	2.6	7.8	{$t$}
\writefig	6.74	0.5	{$U$}
\writefig	10.55	0.5	{$2U$}
\writefig	1.5	5.2	{$-\frac\mu{2d}$}
\writefig	3.75	4.9	{$\frac\mu{2d}$}
\writefig	5.4	3.2	{$-\frac{\mu-U}{4d}$}
\writefig	7.5	3.1	{$\frac{\mu-U}{4d}$}
\writefig	9.3	2.5	{$-\frac{\mu-2U}{6d}$}
\writefig	11.3	2.3	{$\frac{\mu-2U}{6d}$}
\writefig	7.95	5.55	{BEC}
\writefig	1.1	1.9	{\footnotesize $\rho=0$}
\writefig	4.25	1.65	{\footnotesize $\rho=1$}
\writefig	8.3	1.4	{\footnotesize $\rho=2$}
\writefig	11.8	1.3	{\footnotesize $\rho=3$}
}
\caption{Zero temperature phase diagram for the Bose-Hubbard model. Lobes are
incompressible phases with integer densities. Thin lines represent the approximate
critical line defined in \eqref{tcritapprox}.}
\label{figphdBH}
\end{figure}
In order to state our first result, we recall that the pressure $p(\beta,\mu)$ is defined
by
\be
\label{defpressure}
p(\beta,\mu) = \lim_{\Lambda\nearrow\bbZ^d} \frac1{|\Lambda|} \log \Tr_{\caF_\Lambda}
\e{-\beta(H-\mu N)}.
\ee
Here the limit is taken over a sequence of boxes of increasing size; standard arguments ensure its
existence. Its derivative with respect to the chemical potential is the density; i.e.,
\be
\label{defdensity}
\rho(\beta,\mu) = \frac1\beta \frac\partial{\partial\mu} p(\beta,\mu).
\ee
The zero density phase is simpler to analyze because of the absence of quantum
fluctuations. The following theorem holds uniformly in $U$, and therefore also applies to
the hard-core model.

\begin{theorem}[Zero density phase]\hfill
\label{thmzero}

\noindent
For $\mu < -2dt$, there exists $\beta_0$ such that if $\beta>\beta_0$, we have that
\begin{itemize}
\item[(a)] the pressure is real analytic in $\beta,\mu$;
\item[(b)] $\rho(\beta,\mu) < \e{-a\beta}$.
\end{itemize}
Here, $a>0$ depends on $t,\mu,d$, but it is
uniform in $\beta$ and $U$.
\end{theorem}

This theorem is proven in Section \ref{secexpzero}.

The transition lines between the Mott phases of density $\rho\geq1$ and the Bose-Einstein condensate
are much harder to study because of the presence of quantum fluctuations. We consider a
simplified model with a generalized hard-core condition that prevents more than two bosons
from occupying a given site. The Hamiltonian is still given by \eqref{defHam}, but it acts on the
Hilbert space spanned by the configurations $\{0,1,2\}^\Lambda$. The phase diagram of
this model is depicted in Fig.\ \ref{figphdhc2BH}.
This model is the simplest one exhibiting a phase with quantum fluctuations. Notice that,
in the limit $U\to\infty$, this model coincides with the usual hard-core model. The
zero-density phase and the $\rho=2$ phase are characterized by Theorem \ref{thmzero}. The
transition line of the $\rho=1$ phase is more complicated. The following theorem shows
that it is equal to $\mu/2d$ to first order in $t/U$, as in the hard-core model.
\bfig
\epsfxsize=100mm
\centerline{\epsffile{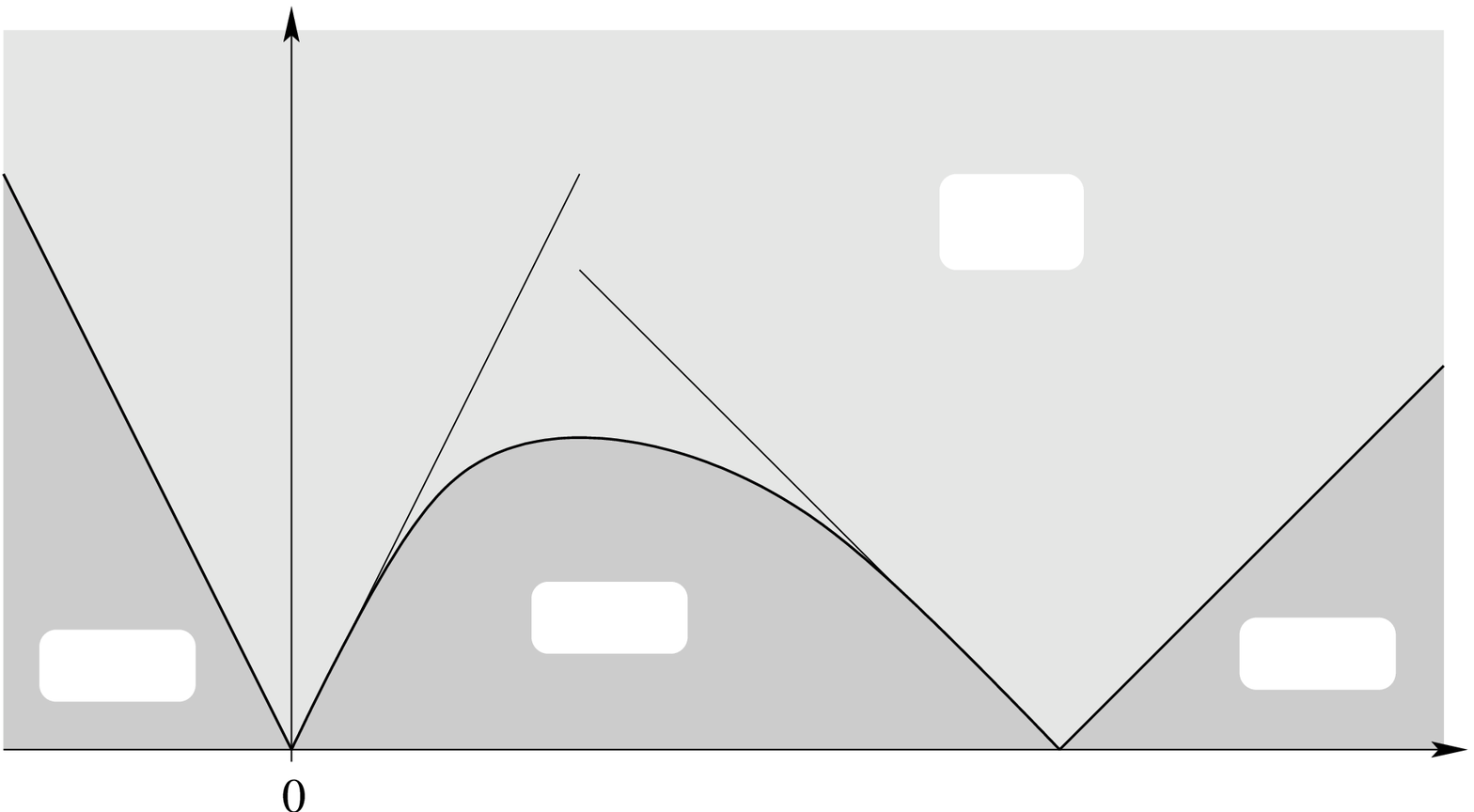}}
\figtext{
\writefig	11.95	0.7	{$\mu$}
\writefig	3.55	5.8	{$t$}
\writefig	8.95	0.5	{$U$}
\writefig	2.5	3.7	{$-\frac\mu{2d}$}
\writefig	4.8	3.9	{$\frac\mu{2d}$}
\writefig	7.0	3.2	{$-\frac{\mu-U}{4d}$}
\writefig	9.9	2.7	{$\frac{\mu-U}{4d}$}
\writefig	8.3	4.3	{BEC}
\writefig	2.3	1.4	{\footnotesize $\rho=0$}
\writefig	5.6	1.7	{\footnotesize $\rho=1$}
\writefig	10.5	1.45	{\footnotesize $\rho=2$}
}
\caption{Zero temperature phase diagram for the Bose-Hubbard model with the generalized
hard-core condition.}
\label{figphdhc2BH}
\end{figure}

\begin{theorem}[Mott phase $\rho=1$ in generalized hard-core model]\hfill
\label{thmmain}

\noindent
Assume that $0<\mu<\frac U4$ and $t < \frac\mu{2d} - {\rm const} \,\frac{t^2}U$ (with ${\rm const}
\leq 2^{11}d$). Then there exist $\beta_0$ and $a>0$ (depending on $d,t,\mu,U$) such
that if $\beta>\beta_0$, we have that
\begin{itemize}
\item[(a)] the pressure is real analytic in $\beta,\mu$;
\item[(b)] $|\rho(\beta,\mu)-1| < \e{-a\beta}$.
\end{itemize}
\end{theorem}

The critical line is expected to be close to $t=\frac\mu{2d}$ for small $t,\mu$, so that our
condition agrees to first order in $t$. While we do not state and prove it explicitly, a
similar claim holds around $4dt = \mu-U$. Indeed, the $\rho=1$ phase prevails for $t
\lesssim \frac{U-\mu}{4t}$ for small $t$. The ``quantum Pirogov-Sinai theory'' of references \cite{BKU,DFF} applies here and allows to establish the existence of a Mott insulator for low $t$. Proving that the domain extends almost to the line $t=\frac\mu{2d}$ requires additional arguments, however; Theorem \ref{thmmain} is proved in Section \ref{secexpMott}. The generalized hard-core condition considerably simplifies the proof. Indeed, it allows for a cute and convenient representation of the grand-canonical partition function in terms of a gas of non-overlapping oriented space-time loops, see Fig.\ \ref{figexcitations} in Section \ref{secexpMott}. The result is nevertheless expected to hold for the regular Bose-Hubbard model as well.

While we cannot establish the presence of a Bose-Einstein condensate, we can prove the absence
of Mott insulating phases away from the critical lines, by establishing bounds on the density
of the system.

\begin{theorem}[Absence of Mott phases]\hfill
\label{thmnoMott}
\begin{itemize}
\item[(a)] For $t>-\frac\mu{2d}$ and for any $\Lambda$ large enough, the density of the ground state is bounded below by a strictly positive constant, that depends on $t,\mu$ but not on $\Lambda$.
This applies to the model with or without hard-core condition.
\item[(b)] Consider the model with generalized hard cores. For $t > \frac\mu{2d} + C t (\frac tU)^{\frac2{d+2}}$ and for any $\Lambda$ large enough, the density of the ground state is less than a constant that is strictly less than 1; it depends on $t,\mu$ but not on $\Lambda$.
\end{itemize}
\end{theorem}

This theorem is proved in Section \ref{secdensbounds}. It is shown that $C \leq \frac{d+2}{2d} (2^{10} d \pi^d)^{\frac2{d+2}}$.

Quantum fluctuations have some influence on the phase diagram, and a
detailed discussion is necessary. ``Quantum fluctuations'' are fluctuations in the ground state
around the constant configuration with $k$ bosons at each site, for some $k$ depending on
$\mu$ and $t$. They are present in Mott phases for
$k\geq1$, while the ground state for $\mu<-2dt$ is simply the empty configuration. Quantum
fluctuations are {\it not} present in effective hard-core models where each site is
allowed
either $k$ or $k+1$ bosons. Their presence lowers the energy of both Mott and Bose condensate
states. The key question is which phase benefits most from them. In other words,
writing the critical hopping parameter as
\be
t_{\rm c}(\mu) = t_{\rm c}^{\rm approx}(\mu) \bigl( 1 + a(\tfrac tU) \bigr),
\ee
the question is about the sign of $a(\frac tU)$, for small $\frac tU$.

The study in \cite{FM}, based on expansion methods (no attempt at a rigorous control of
convergence is made), suggests a rather surprising answer: the sign of $a(\frac tU)$
depends on the dimension! Namely, the quantum fluctuations favor Mott phases for $d=1$, and
they favor the Bose condensate for $d\geq2$. We expect that this question can be rigorously
settled by combining the partial diagonalization method of \cite{DFFR} with our
expansions in Section \ref{secexpMott}.

\section{Low-density expansions}
\label{secexpzero}

In this section, we present a Feynman-Kac expansion of the partition function adapted to the study of quantum states that are perturbations of the zero density phase. In this situation, quantum effects are reduced to a minimum, amounting basically to the combinatorics related to particle indistinguishability.  Nevertheless, the resulting cluster expansion must deal with two difficult points: arbitrarily large
numbers of bosons and closeness to the transition line. Both difficulties are resolved by
estimating the entropy of space-time trajectories in a way inspired by Kennedy's study of the Heisenberg model \cite{Ken} --- the present situation being actually simpler.

The grand-canonical partition function of the Bose-Hubbard model is given by
\be
\label{defpartfct}
Z(\beta,\Lambda,\mu) = \Tr \e{-\beta(H-\mu N)},
\ee
where the trace is taken over the bosonic Fock space. A standard Feynman-Kac
expansion yields an expression for $Z$ in terms of ``space-time trajectories'', i.e.\
continuous-time nearest-neighbor random walks. More precisely,
\bm
\label{FKexp}
Z(\beta,\Lambda,\mu) = \sum_{N\geq0} \frac{\e{\beta\mu N}}{N!}
\sum_{x_1,\dots,x_N\in\Lambda} \sum_{\pi\in S_N} \int\dd\nu_{x_1
x_{\pi(1)}}^\beta(\theta_1) \dots \int\dd\nu_{x_N x_{\pi(N)}}^\beta(\theta_N) \\
\prod_{1\leq
i<j\leq N} \exp\Bigl\{ -U \int_0^\beta \delta_{\theta_i(\tau), \theta_j(\tau)} \dd\tau
\Bigr\}.
\end{multline}
Here, $\theta$ denotes a space-time trajectory, i.e. $\theta$ is a map $[0,\beta] \to
\Lambda$ that is constant except for finitely many ``jumps'' at times
$0<\tau_1<\dots<\tau_m<\beta$, and
$$
|\theta(\tau_j-) - \theta(\tau_j+)| = 1.
$$
The ``measure'' $\nu_{xy}^\beta$ on trajectories starting at $x$ and
ending at $y$ introduced in Eq.\ \eqref{FKexp} is a shortcut for the following operation. If $f$ is a function on trajectories, then
\be
\label{integration}
\int\dd\nu_{xy}^\beta(\theta) f(\theta) = \sum_{m\geq0} t^m
\sumtwo{x_1,\dots,x_{m-1}}{|x_j-x_{j-1}|=1} \int_{0<\tau_1<\dots<\tau_m<\beta} \dd\tau_1
\dots \dd\tau_m f(\theta).
\ee
The second sum is over nearest-neighbor sites such that $|x_1-x|=|x_{m-1}-y|=1$. The
trajectory $\theta$ on the right side of \eqref{integration} is given by
$$
\theta(\tau) = x_j \quad\quad \text{for } j \in [\tau_j,\tau_{j+1}),
$$
where $(x_0,\tau_0) = (x,0)$ and $(x_m,\tau_{m+1}) = (y,\beta)$. The underlying trace operation constraints the ensemble of trajectories to satisfy a periodicity condition in the ``$\beta$-direction".  The  initial and final particle configurations must be identical, modulo particle indistinguishablity.  This explains the sum over permutations of  $N$ elements, $\pi\in S_N$, on the right side of \eqref{FKexp}.

We shall rewrite the expresssion \eqref{FKexp} for the partition function in a form that
fits into the framework of cluster expansions. The main result of cluster expansions is
summarized in the appendix, and it is enough for our purpose.

Trajectories are correlated because of (i) the interactions in the exponential factors of
\eqref{FKexp} which penalize intersections, and (ii) the permutations linking initial and
final sites of different trajectories. The cluster expansion is designed to handle the
former factors, but we need to deal first with the latter issue so to fall into the
required framework.  To this end, we concatenate each original trajectory with the one
starting at its final site, so as to obtain a single closed trajectory that wraps several times around the $\beta$ axis.  Hence, instead of open trajectories $[0,\beta] \to \Lambda$, we consider ensembles of closed trajectories $\theta_i:[0,\ell_i\beta] \to \Lambda$, with $\ell_i$ being their winding number. Each such closed trajectory corresponds to a cycle of length $\ell_i$ of the permutation $\pi$ determined by the endpoints of the component open trajectories. 
For each cycle, the sum over $\ell_i$ sites in $\Lambda$ and the integrals over the $\ell_i$ enchained open trajectories can be written as a sum over a single site $x_i$, followed by an integral over closed trajectories with $\theta_i(0) = \theta(\ell_i \beta) = x_i$. Recalling that there are
$$
\frac1{k!} \frac{N!}{\prod_{i=1}^k \ell_i}
$$
permutations with $k$ cycles of lengths $\ell_1,\dots,\ell_k$, we obtain the following expansion of the partition function in terms of closed trajectories instead of particles:
\bm
\label{cycleexp}
Z(\beta,\Lambda,\mu) = \sum_{k\geq0} \frac1{k!} \sum_{x_1,\dots,x_k\in\Lambda}
\sum_{\ell_1,\dots,\ell_k\geq1} \int\dd\nu_{x_1 x_1}^{\ell_1 \beta}(\theta_1) \dots
\int\dd\nu_{x_k x_k}^{\ell_k \beta}(\theta_k) \\
\prod_{i=1}^k w(\theta_i) \prod_{1\leq i<j\leq k} \bigl( 1 - \zeta_U(\theta_i,\theta_j)
\bigr).
\end{multline}
Let $\ell(\theta)$ denote the winding number of the trajectory $\theta : [0,\ell(\theta)
\beta] \to \Lambda$. Its weight is defined by
\be
\label{defweighttraj}
w(\theta) = \tfrac1{\ell(\theta)} \e{\beta\mu\ell(\theta)} \exp\bigl\{ -U W(\theta) \bigr\}.
\ee
Here, $W(\theta)$ measures the self-intersection of $\theta$, that is,
\be
W(\theta) = \sum_{0\leq i<j\leq\ell(\theta)-1} \int_0^\beta
\delta_{\theta(i\beta+\tau),\theta(j\beta+\tau)} \dd\tau.
\ee
It will suffice to use the bound $w(\theta) \leq \frac1{\ell(\theta)}
\e{\beta\mu\ell(\theta)}$. Finally,
interactions between trajectories $\theta$ and $\theta'$ are given by
\be
\zeta_U(\theta,\theta') = 1 - \exp\bigl\{ -U W(\theta,\theta') \bigr\}.
\ee
Here, $W(\theta,\theta')$ measures the overlap between trajectories $\theta$ and
$\theta'$,
\be
W(\theta,\theta') = \sum_{i=0}^{\ell(\theta)-1} \sum_{j=0}^{\ell(\theta')-1}
\int_0^\beta \delta_{\theta(i\beta+\tau),\theta'(j\beta+\tau)} \dd\tau.
\ee
Expression \eqref{cycleexp} is suited for an application of Theorem \ref{thmclexp}. We
show that the weights $w(\theta)$ are small in the sense that they satisfy the
``Koteck\'y-Preiss criterion'' \eqref{KPcrit}.

\begin{proposition}
\label{propKPcritzero}  For each closed trajectory $\theta$ let $j(\theta)$ denote the number of jumps of $\theta$.  Then, there exist constants $a,b>0$ such that 
$$
\sum_{\ell'\geq1} \frac{\e{\beta\mu\ell'}}{\ell'} \sum_{x\in\Lambda}
\int\dd\nu_{xx}^{\ell'\beta}(\theta') \e{a j(\theta') + \beta b \ell'}
\zeta_U(\theta,\theta') \leq a j(\theta) + \beta b \ell(\theta).
$$
\end{proposition}

\begin{proof}
Since $\zeta_U(\theta,\theta')$ is increasing in $U$, it is enough to prove that, for any trajectory $\theta$,
\be
\label{suffisant}
\sum_{\ell'\geq1}\frac {\e{\beta(\mu+b)\ell'}}{ \ell'} \sum_x \int\dd\nu_{xx}^{\ell'\beta}(\theta')\, \e{a
j(\theta')} \, \zeta_\infty(\theta,\theta')
\;\leq\; a j(\theta) + \beta b \ell(\theta)\;,
\ee
with
\be
\begin{split}
\zeta_\infty(\theta,\theta') &= \upchi \Bigl[ \theta(i\beta+\tau) =
\theta'(k\beta+\tau) \text{ for some } 0\leq i\leq\ell(\theta)-1, \\
&\hspace{60mm} 0\leq k\leq\ell(\theta')-1, 0<\tau<\beta \Bigr] \;.
\end{split}
\nn
\ee
Here, $\upchi[\cdot]$ denotes the characteristic function of the event in brackets.

A trajectory $\theta'$ intersects $\theta$ if a jump of $\theta'$ intersects a vertical
line of $\theta$, or if a jump of $\theta$ intersects a vertical line of $\theta'$ (or
both). Let $\nu_0^\beta$ denote the measure on trajectories $[0,\beta] \to \bbZ^d$, starting at $x=0$
and with a jump at $\tau=0$. Integration with respect to $\nu_0^\beta$ can be
defined similarly as in \eqref{integration}; formally, we can also write
\be
\int\dd\nu_0^\beta(\theta) f(\theta) = \int\dd\nu_{00}^\beta(\theta) \delta(\tau_1)
f(\theta),
\ee
where $\nu_{00}^\beta$ is as in \eqref{integration} (with $x=y=0$), and where
the Dirac function $\delta(\tau_1)$ forces the first jump to occur at $\tau_1=0$.
We get an upper bound by neglecting the restriction that trajectories need to remain in
$\Lambda$. The left side of \eqref{suffisant} is then bounded by
\be
\label{cestuneborne}
j(\theta) \sum_{\ell'\geq1} \frac{\e{\beta (\mu+b) \ell'}}{\ell'} \int\dd\nu_{00}^{\ell'\beta} \e{a
j(\theta')} + \beta \ell(\theta) \sum_{\ell'\geq1} \frac{\e{\beta (\mu+b) \ell'}}{\ell'}
\int\dd\nu_0^{\ell'\beta} \e{a j(\theta')}.
\ee
The first term accounts for trajectories $\theta'$ intersecting jumps of $\theta$; the
second term accounts for trajectories $\theta'$ involving a jump that intersects a
vertical line of $\theta$. We integrate over all trajectories $[0,\ell'\beta] \to \bbZ^d$ that start at $x=0$,
without requiring them to stay in $\Lambda$.

Each trajectory $\theta'$ in the last integral of \eqref{cestuneborne} can be decomposed into the 
jump from a neighbor $z$ into $0$, which contributes a factor $t\e{a}$ (see definition
\eqref{integration}), plus a trajectory from $z'$ to $0$.  As $0$ has $2d$ neighbors we see that
\be
\label{r.10}
\int\dd\nu_{0}^{\ell'\beta}(\theta')\,\e{aj(\theta')} \leq 2dt\e{a}\,
\int\dd\nu_{z0}^{\ell'\beta}(\theta')\,\e{aj(\theta')}\;.
\ee
Furthermore, the definition  \eqref{integration} implies that for every $x,y$,
\be
\label{r.5}
\begin{split}
\int\dd\nu_{xy}^{\ell'\beta}(\theta')\,\e{aj(\theta')} &=
\sum_{m\geq0} (t\e{a})^m
\sumthree{x_1,\dots,x_{m-1}}{|x_j-x_{j-1}|=1}{|x_1-x|=|y-x_{m-1}|=1} 
\int_{0<\tau_1<\dots<\tau_m<\ell'\beta} \dd\tau_1\dots \dd\tau_m \\
& \le\; \e{t\e{a}2d\ell'\beta}\;.
\end{split}
\ee

From \eqref{suffisant}, \eqref{cestuneborne}, \eqref{r.10} and \eqref{r.5} we conclude that
\be
\label{r.6}
\begin{split}
& \sum_{\ell'\geq1}\frac {\e{\beta(\mu+b)\ell'}}{ \ell'} \sum_x \int\dd\nu_{xx}^{\ell'\beta}(\theta')\, \e{a
j(\theta')} \, \zeta_\infty(\theta,\theta')\\
&\hspace{10mm} \le 
\bigl(j(\theta)+2dt\e{a}\beta\ell(\theta)\bigr) \sum_{\ell'\geq1} \exp\bigl\{ \beta\ell' [\mu + 2dt \e a + b] \bigr\}.
\end{split}
\ee
As $\mu + 2dt < 0$, we can choose $a$ and $b$ such that $\mu + 2dt \e a + b < 0$. Then
\eqref{suffisant} holds for $\beta$ large enough.
\end{proof}

\begin{proof}[Proof of Theorem \ref{thmzero}]
Recall expression \eqref{defpressure} for the pressure. Proposition
\ref{propKPcritzero} establishes the convergence of cluster expansions, as stated in
Theorem \ref{thmclexp}. With $\varphi$ denoting the usual combinatorial function of
cluster expansions, see \eqref{defphi}, the partition function has the
absolutely convergent expression
\bm
Z(\beta,\Lambda,\mu) = \exp\Bigl\{ \sum_{m\geq1} \sum_{x_1,\dots,x_m\in\Lambda}
\sum_{\ell_1,\dots,\ell_m\geq1} \int\dd\nu_{x_1 x_1}^{\ell_1\beta}(\theta_1) \\
\dots \int\dd\nu_{x_m x_m}^{\ell_m\beta}(\theta_m) \, \varphi(\theta_1,\dots,\theta_m)
\prod_{j=1}^m w(\theta_j) \Bigr\}.
\end{multline}
Taking the logarithm and dividing by the volume, standard arguments show that boundary
terms vanish in the thermodynamic limit, and we obtain
\bm
p(\beta,\mu) = \sum_{m\geq1} \sum_{x_2,\dots,x_m\in\bbZ^d} \sum_{\ell_1,\dots,\ell_m\geq1}
\int\dd\nu_{00}^{\ell_1\beta}(\theta_1) \int\dd\nu_{x_2 x_2}^{\ell_2\beta}(\theta_2) \\
\dots \int\dd\nu_{x_m x_m}^{\ell_m\beta}(\theta_m) \, \varphi(\theta_1,\dots,\theta_m)
\prod_{j=1}^m w(\theta_j).
\end{multline}
Integrals can be viewed as functions of $\beta,\mu$, indexed by $m$, $(x_i)$, and
$(\ell_i)$. They are real analytic in the domain $(\beta,\mu): \beta > \beta_0(\mu)$.
Their sum is absolutely convergent and Vitali's convergence theorem implies that
$p(\beta,\mu)$ is analytic.

Recall that the density is given by the derivative of the pressure with respect to the
chemical potential; see \eqref{defdensity} for the precise definition. The analyticity implied by the expansion allows for term-by-term differentiation.  We can check that
\bm
\label{expression_densite}
\rho(\beta,\mu) = \sum_{\ell_1\geq1} \int\dd\nu_{00}^{\ell_1\beta}(\theta_1)
\frac{\partial w(\theta_1)}{\partial\mu} \sum_{m\geq1} m \sum_{x_2,\dots,x_m\in\bbZ^d}
\sum_{\ell_2,\dots,\ell_m\geq1} \\
\int\dd\nu_{x_2 x_2}^{\ell_2\beta}(\theta_2) \dots
\int\dd\nu_{x_m x_m}^{\ell_m\beta}(\theta_m) \, \varphi(\theta_1,\dots,\theta_m)
\prod_{j=2}^m w(\theta_j).
\end{multline}
Note that $\frac\partial{\partial\mu} w(\theta) = \beta \ell(\theta) w(\theta)$, as follows from
definition \eqref{defweighttraj} of the weight of trajectories. By \eqref{boundclusters}, we have the bound
\be
\rho(\beta,\mu) \leq \beta \sum_{\ell_1\geq1} \ell_1 \int\dd\nu_{00}^{\ell_1\beta}(\theta)
w(\theta) \e{a(\theta)}.
\ee
There exists $\varepsilon>0$ such that $\mu + 2dt \e a + b + \varepsilon < 0$. Using
\eqref{r.5}, we get
\be
\rho(\beta,\mu) \leq \beta \e{-\varepsilon\beta} \sum_{\ell_1\geq1} \ell_1 \e{\beta\ell_1 [\mu +
2dt\e a + b + \varepsilon]}.
\ee
Then $\rho(\beta,\mu) \leq \e{-\varepsilon\beta}$ for $\beta$ large enough, and this
completes the proof of Theorem \ref{thmzero}.
\end{proof}

\section{Space-time loop representation}
\label{secexpMott}

The study of the transition line for the Mott phase with unit density requires the
analysis of perturbations of the ``vacuum'' formed by one particle at each site.  This
involves the control of full-fledged quantum fluctuations.  We turn, then, to a more
general expansion setting previously employed to study spin and fermionic systems
\cite{BKU,DFF}.  This setting shares some similarities with that of Section \ref{secexpzero}, but it also differs from it in significant ways. We use the same symbols $\nu, w, \zeta, \ell, \theta$, but we caution the reader that they are defined in slightly different ways. Besides the quantum-fluctuation issue, bosonic systems present the additional complication of the unboundedness of occupation numbers.  In the present paper we wish to leave this second issue aside. We consider, thus, the model with generalized hard-core condition that ensures that configurations have at most two bosons at each site.

Recall definition \eqref{defpartfct} of the grand-canonical partition function.
It is convenient to write
\be
H - \mu N = V + T,
\ee
where $V$ denotes the diagonal terms (i.e., interactions and chemical potential terms) in the basis of occupation numbers in position
space, and $T$ denotes the hopping
terms. We will consider $T$ to be a perturbation of $V$. Our expansion is based on Duhamel's
formula,
\be
\e{-\beta (V+T)} = \e{-\beta V} + \int_0^\beta \dd\tau \, \e{-\tau V} (-T) \,\e{-(\beta-\tau)
(V+T)},
\ee
which we can iterate to obtain
\be
\e{-\beta (V+T)} = \sum_{m\geq0} \int_{0<\tau_1<...<\tau_m<\beta} \dd\tau_1 \dots \dd\tau_m
\, \e{-\tau_1 V} (-T) \e{-(\tau_2-\tau_1) V} \dots (-T) \e{-(\beta-\tau_m) V}.
\ee
Then
\bm
Z(\beta,\Lambda,\mu) = \Tr \e{-\beta (V+T)} = \sum_{m\geq0} t^m
\int_{0<\tau_1<...<\tau_m<\beta} \dd\tau_1 \dots \dd\tau_m \\
\sum_{(x_1,y_1), \dots, (x_m,y_m)} \Tr \e{-\tau_1 V} c_{x_1}^\dagger c_{y_1}
\e{-(\tau_2-\tau_1) V} \dots c^\dagger_{x_m} c_{y_m} \e{-(\beta-\tau_m) V}.
\label{expansion}
\end{multline}

We denote by $n = (n_x)_{x\in\Lambda}$, $n_x \in \bbN$, a ``classical configuration'' that
represents the state where $n_x$ bosons are located at site $x$, and $\ket n$ the
corresponding normalized vector.  Inserting projector decompositions ${\bf 1}=\sum_{n_i} \ket{n_i}\bra{n_i}$ the trace can be written as
\be
\label{r.7}
\begin{split}
& \Tr \e{-\tau_1 V} c_{x_1}^\dagger c_{y_1}
\e{-(\tau_2-\tau_1) V} \dots c^\dagger_{x_m} c_{y_m} \e{-(\beta-\tau_m) V}\;=\\
&\hspace{5mm}  \sum_{n_0,n_1,\dots,n_m}
\bra{n_0} \e{-\tau_1 V} c_{x_1}^\dagger c_{y_1}\ket{n_1}\bra{n_1}
\e{-(\tau_2-\tau_1) V} \dots c^\dagger_{x_m} c_{y_m} \ket{n_m} \\
& \hspace{35mm} \times\ 
\bra{n_m}\e{-(\beta-\tau_m) V}\ket{n_0}\;.
\end{split}
\ee

As the operator $V$ is diagonal in the base $\ket n$, this decomposition allows us to rewrite the expansion \eqref{expansion} in the form
\be
\label{expquconf}
Z(\beta,\Lambda,\mu) = \int\dd\nu(\bsn)\, w(\bsn)\;,
\ee
where
\begin{itemize}\item[(i)]
$\bsn$ is a  space-time {\it quantum configuration}, namely an assignment of a configuration $\bsn(\tau)$, for each $0<\tau<\beta$, such that
\begin{itemize}
\item $\bsn$ is constant in $\tau$, except at finitely many times $\tau_1<...<\tau_m$,
with $m$ even.
\item At each $\tau_i$, a ``jump'' occurs, i.e.\ there are nearest-neighbor sites
$(x_i,y_i)$ such that
\be
\label{defjump}
\bsn_x(\tau_i+) = \begin{cases} \bsn_x(\tau_i-)+1 & \text{if } x=x_i, \\ \bsn_x(\tau_i-)-1 & \text{if } x=y_i, \\ \bsn_x(\tau_i-) & \text{otherwise.} \end{cases}
\ee
\item $\bsn$ is periodic in the $\tau$ direction: $\bsn(\beta)=\bsn(0)$.
\end{itemize}

\item[(ii)]
 $w(\bsn)$ are positive weights defined by
\be
\label{weightquantconf}
w(\bsn) = \exp\Bigl\{ -\int_0^\beta V(\bsn(\tau)) \,\dd\tau \Bigr\} \prod_{i=1}^m \Bigl[ t \sqrt{\bsn_{x_i}(\tau_i+) \bsn_{y_i}(\tau_i-)} \Bigr],
\ee
with the short-hand notation $V(n) \equiv \bra n V \ket n$. 

\item[(iii)]  Integration with respect to the ``measure'' $\nu$ on quantum configurations
stands for a sum over configurations at time 0, a sum over $m$, integrals over jumping times, and
sums over locations of jumps. 
\end{itemize}

The expansion just obtained is rather general.  It is convenient to interpret it in terms
of random geometrical objects in a model-dependent fashion.  For the case of interest
here, we follow the ``excitations'', namely the sites where the occupation number is
different from the vacuum value 1.  We therefore embed  the ``space time'' $\Lambda \times [0,\beta]$ in the cylinder
$\bbR^d \times S^1$ (with periodic boundary conditions in the time direction) and decompose the trajectories of the excitations in connected components.  In this way, a quantum configuration $\bsn$ can be represented as a set of
\emph{non-intersecting} loops (with winding numbers $n = 0, \pm1, \pm2, \dots$) in this cylinder.  The representation is defined by the following rules:
\begin{itemize}
\item The constant configuration $\bsn$ with $n_x(\tau)=1$, for all $x\in\Lambda$ and
$0\leq\tau\leq\beta$, has no loops.
\item A jump of a boson from $y_i$ to $x_i$ at time $\tau_i$ (see \eqref{defjump}) is
represented by a horizontal arrow from $(y_i,\tau_i)$ to ($x_i,\tau_i)$.
\item The points $(x,\tau)$ with $n_x(\tau)\neq1$ are represented by vertical segments.
These segments point upwards if $n_x(\tau)=2$, and downwards if $n_x(\tau)=0$.
\end{itemize}
Loops are illustrated in Fig.\ \ref{figexcitations}. Similar representations have been
used in various contexts, e.g.\ in a study of the Falicov-Kimball model \cite{MM}.
\bfig
\epsfxsize=80mm
\centerline{\epsffile{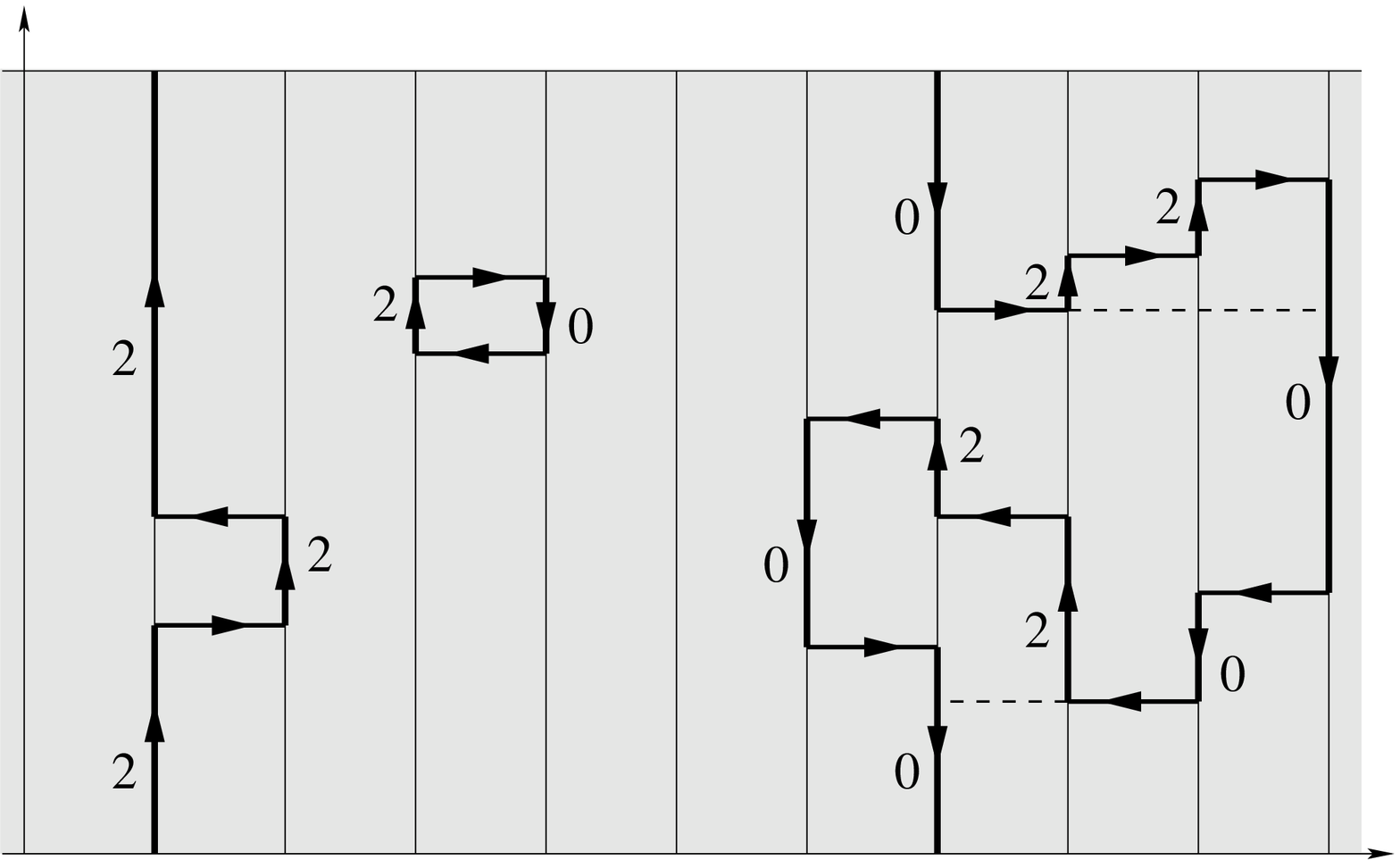}}
\figtext{
\writefig	11.0	0.4	{$\Lambda$}
\writefig	2.5	4.9	{$\beta$}
}
\caption{Illustration for the gas of space-time loops. There are three loops with respective
winding numbers 1,0, and -1.}
\label{figexcitations}
\end{figure}
Given a loop $\gamma$, we introduce the number of jumps $j(\gamma)$ (always an even
number, possibly zero); the length $\ell_0(\gamma)$ of all vertical segments pointing
downwards; the length $\ell_2(\gamma)$ of all vertical segments pointing upwards;
$\ell(\gamma) = \ell_0(\gamma)+\ell_2(\gamma)$; and the winding number $z(\gamma)$.
Notice that $\ell_2(\gamma)-\ell_0(\gamma)=\beta z(\gamma)$. A
loop $\gamma$ defines a unique quantum configuration $\bsn^\gamma$. We define the {\it weight} of a
loop as
\be
\label{defweightloop}
w(\gamma) = t^{j(\gamma)} \prod_{i=1}^{j(\gamma)} \sqrt{\bsn_{y_i}^\gamma(\tau_i-)
\bsn_{x_i}^\gamma(\tau_i+)} \; \e{-\ell_0(\gamma) \mu} \e{-\ell_2(\gamma) (U-\mu)}.
\ee
Note that we have subtracted the classical energy of the background configuration with one
boson at each site. The weight $w(\gamma)$ thus only depends on excitation energies.

These definitions allow us to rewrite the partition function \eqref{expquconf} in terms of
loops and their weights instead of space-time configurations. Unlike the trajectories of Section \ref{secexpzero}, the loops here have only a hard-core interaction due to the requirement of non-intersection.  Furthermore, if
$\Gamma = \{\gamma_1,\dots,\gamma_m\}$ is a set of disjoint loops,
we have the important property that the weight of the corresponding quantum configuration $\bsn^\Gamma$ factorizes,
\be
w(\bsn^\Gamma) = \e{\beta\mu|\Lambda|} \prod_{i=1}^m w(\gamma_i).
\ee
We define a measure on loops, also denoted $\nu$, and we rewrite the
partition function as
\be
\label{partfctexpanded}
Z(\beta,\Lambda,\mu) = \e{\beta\mu|\Lambda|} \sum_{m\geq0} \frac1{m!}
\int\dd\nu(\gamma_1) \dots \int\dd\nu(\gamma_m) \prod_{i=1}^m w(\gamma_i)
\prod_{1\leq i<j\leq m} \bigl( 1 - \zeta(\gamma_i,\gamma_j) \bigr).
\ee
Here, the term corresponding to $m=0$ is set to $\e{\beta\mu|\Lambda|}$, and the function
$\zeta(\gamma,\gamma')$ equals 1 if the loops $\gamma$ and $\gamma'$ intersect (more precisely, if some of their
vertical segments intersect), and equals 0 if the loops have disjoint support.

The expression \eqref{partfctexpanded} for the partition function is an adequate starting
point for the method of cluster expansions. We prove that the weights are small so as to
satisfy the ``Koteck\'y-Preiss criterion'', Eq.\ \eqref{KPcrit}. We can then
appeal to \ Theorem \ref{thmclexp} to conclude that the cluster expansion
converges.

\begin{proposition}
\label{propKPcrit}
Under the hypotheses of Theorem \ref{thmmain}, we have that, for any loop $\gamma$,
$$
\int \dd\nu(\gamma') \,w(\gamma')\, \zeta(\gamma,\gamma')\, \e{a(\gamma')}
\; \leq\; a(\gamma)
$$
with
\be
a(\gamma) = 2^{14}d \tfrac{t^2}{U^2} j(\gamma) + 2^{12}d \tfrac{t^2}U \ell(\gamma).
\ee
\end{proposition}

Its proof relies on the bounds stated in the following lemma.
Let us partition the set of loops into $\caL = \caL^{(0+)} \cup \caL^{(-)}$, where
$\caL^{(0+)}$ (resp.\ $\caL^{(-)}$) is the set of loops with nonnegative (resp.\ negative)
winding numbers. For each site $z$ we introduce the measures $\nu_z$ on loops
that make a jump at time $\tau=0$ involving $z$. Further, we let
$\caL_z$, $\caL^{(0+)}_z$, and $\caL_z^{(-)}$ denote the sets of loops that contain $(z,0)$.

\begin{lemma}
\label{lemKPcrit}
Under the hypotheses of Theorem \ref{thmmain}, for any site $z$,
\begin{itemize}
\item[(a)] $\int_{\caL^{(0+)}} \dd\nu_z(\gamma)\, w(\gamma)\, \e{a(\gamma)} \;\leq\;
2^{11} d \tfrac{t^2}U.$
\item[(b)] $\int_{\caL^{(0+)}_z} \dd\nu(\gamma) \;w(\gamma) \;\e{a(\gamma)}\; \leq\; \e{-\beta
(U-\mu-2^{12}d \frac{t^2}U)} + 2^{13} d \frac{t^2}{U^2}$.
\item[(c)] $\int_{\caL^{(-)}} \dd\nu_z(\gamma)\, w(\gamma)\, \e{a(\gamma)} \;\leq\; 4dt \,
\e{-\beta(\mu - 2dt - 2^{12}d^2 \frac{t^2}U)}$. 
\item[(d)] $\int_{\caL^{(-)}_z} \dd\nu(\gamma)\, w(\gamma)\, \e{a(\gamma)}
\;\leq\;  \e{-\beta (\mu - 2dt - 2^{12}d^2 \frac{t^2}U)}$. 
\end{itemize}
\end{lemma}

\begin{proof}[Proof of Proposition \ref{propKPcrit}]
Suppose that the loops $\gamma$ and $\gamma'$ intersect, i.e.\ $\zeta(\gamma,\gamma')=1$.
Then either a jump of $\gamma'$ intersects a vertical line of $\gamma$, or a jump of
$\gamma$ intersects a vertical line of $\gamma'$
(both may happen at the same time). The first situation is analyzed using the measures
$\nu_z$, and the second situation involves the sets $\caL_z^{(0+)}$ and $\caL_z^{(-)}$.
More precisely, we have that
\be
\begin{split}
& \int\dd\nu(\gamma')\, w(\gamma') \,\zeta(\gamma,\gamma')\, \e{a(\gamma')}\; \leq\\
& \hspace{15mm} \ell (\gamma) \,\sup_z\int_{\caL} \dd\nu_z(\gamma')\, w(\gamma') \,\e{a(\gamma')} + j(\gamma)
\sup_z \int_{\caL_z} \dd\nu(\gamma')\, w(\gamma')\, \e{a(\gamma')}\;.
\end{split}
\ee
Using the estimates in Lemma \ref{lemKPcrit}, the right side is seen to be smaller than
$a(\gamma)$, provided $\beta$ is large enough.
\end{proof}

\begin{proof}[Proof of Lemma \ref{lemKPcrit}, (a) and (b)]
Loops of $\caL^{(0+)}$ have large energy cost, so crude entropy estimates are enough.
Since $\ell_2(\gamma) \geq \ell_0(\gamma)$ for any loop $\gamma \in
\caL^{(0+)}$, we have that $\mu\ell_0(\gamma) + (U-\mu)\ell_2(\gamma) \geq \frac12
U\ell(\gamma)$. Then
\be
\mu \ell_0(\gamma) + (U-\mu) \ell_2(\gamma) - 2^{12}d \tfrac{t^2}U \ell(\gamma) \;\geq\;
\tfrac14 U \ell(\gamma).
\ee
Further, we can check that 
\be
\label{r.11}
\e{2^{14}d t^2/U^2}\; <\; 2\;.
\ee
From these observations and \eqref{defweightloop}, we obtain that
\be
\label{r.12}
w(\gamma) \e{a(\gamma)}\; \leq\; \left\{\begin{array}{ll}
\e{-\beta (U-\mu-2^{12}d \frac{t^2}U)} & \text{if } j=0\\[5pt]
(4t)^{j(\gamma)}\, \e{-\frac14 U\ell(\gamma)} & \text{if } j\geq 2.
\end{array}\right.
\ee

A loop with $j(\gamma)=2n$ is characterized by a sequence of jump times $0\leq
\tau_1<\tau_2<\cdots<\tau_{2n}$.  At each such time the trajectory can choose among at
most $2d$ neighbors to jump to and 2 directions of time to proceed after the jump. The
last jump is determined by the fact that $\gamma$ must be a loop, so there is no factor
$2d$ (but both time directions are possible). The measure $\nu_z$ involves only loops with
two jumps or more. From the last bound in \eqref{r.12} we obtain
\be
\label{r.13}
\begin{split}
\int_{\caL^{(0+)}} \dd\nu_z(\gamma)\, w(\gamma) \,\e{a(\gamma)}
\;&\leq\;\sum_{n\geq 1} 2 \cdot 2^{2n} (2d)^{2n-1} (4t)^{2n} \biggl(
\int_0^{\infty} d\tau \,\e{-\frac14 U\tau} \biggr)^{2n-1} \\
& = \frac{2^{10} d t^2 / U}{1 - (\frac{2^6 dt}U )^2} \leq 2^{11} d \tfrac{t^2}U.
\end{split}
\ee 

Part (b) of the lemma follows from \eqref{r.12} and from considerations similar to
\eqref{r.13}. Namely,
\bm
\int_{\caL^{(0+)}_z} \dd\nu(\gamma)\, w(\gamma) \,\e{a(\gamma)}
\;\leq\; \e{-\beta (U-\mu-2^{12}d \frac{t^2}U)} \\
+ \sum_{n\geq 1} 2 \cdot 2^{2n} (2d)^{2n-1} (4t)^{2n} \biggl(
\int_0^{\infty} d\tau \,\e{-\frac14 U\tau} \biggr)^{2n}.
\end{multline}
The first term in the right side represents loops without jumps.
The right side is less than the upper bound in Lemma \ref{lemKPcrit} (b).
\end{proof}

\begin{proof}[Proof of Lemma \ref{lemKPcrit}, (c) and (d)]

Loops of $\caL^{(-)}$ have small energy cost when parameters are close to the transition
line. Estimates are needed that are more subtle than for loops of $\caL^{(0+)}$. The
situation is similar to that of Section \ref{secexpzero}, but a problem needs to be solved: Loops,
unlike trajectories, can backtrack in time. Our strategy is to first
``renormalize'' a loop $\gamma \in \caL^{(-)}$ by identifying a trajectory
$\theta=\theta(\gamma)$ that moves only downwards, but with arbitrarily long jumps.
Contributions of backtracking can be controlled by similar estimates as above. The
entropy of these trajectories can be expressed using an appropriate hopping matrix and we obtain sharp enough bounds.

We start with (d). Given a loop $\gamma \in \caL^{(-)}_0$, we start at
$(x,\tau)=(z,\beta)$ and move downwards along $\gamma$. When reaching the end of a vertical
segment (because of the presence of a nearest-neighbor jump), we ignore possible
backtracking and directly jump to the next downwards vertical segment in the loop, at
constant time. See the dotted lines in Fig.\ \ref{figexcitations}. We obtain a trajectory,
since the motion is downwards only, punctuated by with long-range hoppings with which we must cope.

Behind a hop  from $x$ to $y$ there is a backtracking excursion between these sites.  Its
contribution to the total weight of the original loop (times $\e{a(\gamma)}$) is given
by the ``hopping matrix'' component
\be
\label{defsigma}
\sigma_{xy} = \int_{x\to y} \dd\nu_x(\gamma) w(\gamma) \e{a(\gamma)},
\ee
where the integral is over loops that are open, have nonnegative winding number, start with a jump
at $(x,0)$, and end at $(y,0)$. 

Each trajectory so constructed is characterized by a sequence of hopping times $0=\tau_1<\cdots <\tau_{2m}\le\beta$ and a sequence of not-necessarily neighboring sites $x=x_0,x_1,\cdots,  x_{m}=x$ which are the successive hopping endpoints. Its weights are determined by factors exponentally decreasing with $\ell_0$ for each vertical segment and hopping matrix entries for each jump.  In this way we obtain
\be
\label{encoreuneborne}
\begin{split}
\int_{\caL_z^{(-)}} \dd\nu(\gamma) \,w(\gamma)\, \e{a(\gamma)} \;&\leq\;\e{-\beta (\mu - 2^{12}d \frac{t^2}U)}
\sum_{m\geq 0} \frac{\beta^m}{m!} \sumtwo{x_1,\cdots , x_{m-1}}{x_0=x_m=z}\prod_{i=1}^m
\sigma_{x_ix_{i-1}} \\
&\leq\; \e{-\beta (\mu - 2^{12}d \frac{t^2}U)} \; \e{\beta \sum_{x\neq0} \sigma_{0x}}.
\end{split}
\ee
The overall exponential factor comes from the fact that $\ell_0\geq\beta$ because the winding number of the loops is not zero.  The factor $\beta^m/m!$ follows by integrating all choices of hopping times.

To conclude, we must bound the sum of the matrix elements of $\sigma$. The contribution of
open loops that consist in just one jump is $2dt \e{2^{14}d \frac{t^2}{U^2}}$. Other open loops involve two jumps or
more. Each jump has $2d$ possible directions. There are two possible time directions after
each jump, except for the first and last ones. We need to integrate over time occurrence
for each jump except the first one. We obtain
\be
\label{uneautreborne}
\begin{split}
\sum_{x\neq0} \sigma_{0x}\; &\leq\; 2dt\, \e{2^{14}d \frac{t^2}{U^2}} + \sum_{m\geq2}
(2d)^m 2^{m-2} (4t)^m \biggl( \int_0^\infty \e{-\frac14 U\tau} \dd\tau \biggr)^{m-1} \\
&\leq\; 2dt\, \e{2^{14}d \frac{t^2}{U^2}} + 2^9 d^2 \tfrac{t^2}U.
\end{split}
\ee
We used \eqref{r.11}. Inserting into \eqref{encoreuneborne} we obtain Lemma
\ref{lemKPcrit} (d).  The bound of part (c) is similar, with an extra factor $2dt \e{2^{14}d
\frac{t^2}{U^2}}\leq 4dt$ for the additional first jump.
\end{proof}

\begin{proof}[Proof of Theorem \ref{thmmain}]
This proof is similar to the one of Theorem \ref{thmzero}. We use cluster expansions, in order to
get a convergent expansion for the pressure, and prove analyticity by using Vitali's theorem. The
density has an expansion reminiscent of \eqref{expression_densite}, namely
\bm
\rho(\beta,\mu) = 1 + \int_{\caL_0} \dd\nu(\gamma_1) \frac{\partial
w(\gamma_1)}{\partial\mu} \\
\sum_{m\geq1} m \int \dd\nu(\gamma_2) \dots
\int \dd\nu(\gamma_m) \varphi(\gamma_1,\dots,\gamma_m) \prod_{i=2}^m w(\gamma_j).
\end{multline}
The combinatorial function function $\varphi$ is given by \eqref{defphi}.
From \eqref{defweightloop}
\be
\frac{\partial w(\gamma_1)}{\partial\mu} = [\ell_2(\gamma) - \ell_0(\gamma)] w(\gamma).
\ee
Again using Eq.\ \eqref{boundclusters}, we find the bound
\be
|\rho(\beta,\mu) - 1| \leq \int_{\caL_0} \dd\nu(\gamma) \, \bigl| \ell_2(\gamma) - \ell_0(\gamma)
\bigr| \, w(\gamma) \e{a(\gamma)}.
\ee
Only loops with nonzero winding number contribute. Going over  the proof of Lemma
\ref{lemKPcrit} (b) and (d) with $a(\gamma)\to a(\gamma)+\log\ell(\gamma)$, we can check that the right side of the equation above is
less than $\e{-\varepsilon\beta}$ whenever $\mu - 2dt - 2^{12}d^2\frac{t^2}U - \varepsilon >
0$ and $\beta$ is large enough.
\end{proof}

\section{Density bounds}
\label{secdensbounds}

\begin{proof}[Proof of Theorem \ref{thmnoMott}, (a)]
The Bose-Hubbard Hamiltonian preserves the total number of particles, so that the
density can be fixed. We denote by $e_0(\rho)$ the ground state
energy per site in the subspace of density $\rho$. Neglecting repulsive interactions can only
decrease the ground state energy; the minimum kinetic energy of a single boson is $-2dt$. It
follows that $e_0(\rho) \geq (-\mu-2dt) \rho$ for all $U\geq0$.

\bfig
\epsfxsize=80mm
\centerline{\epsffile{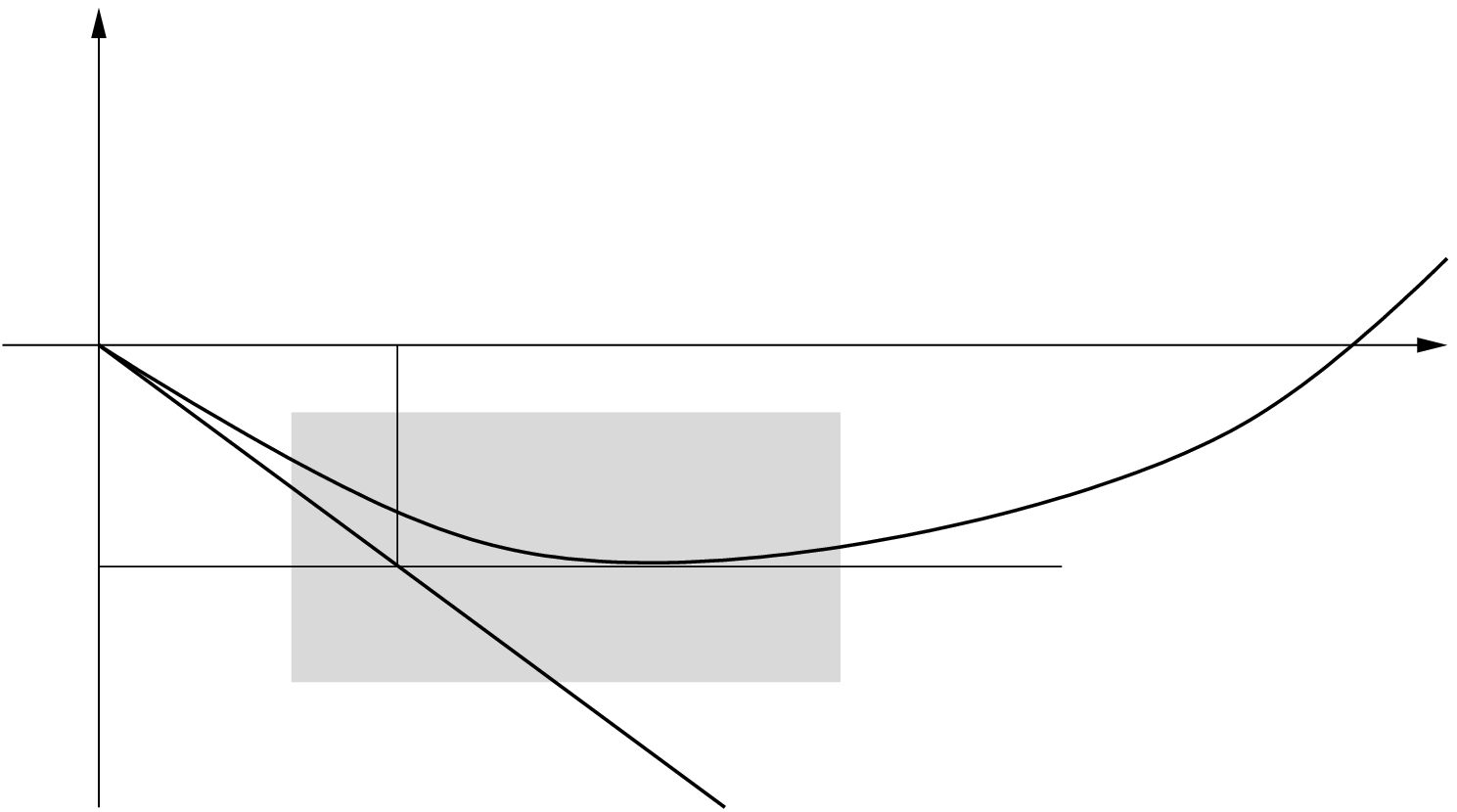}}
\figtext{
\writefig	10.7	2.8	{\footnotesize $\rho$}
\writefig	3.6	4.7	{\footnotesize $e(\rho)$}
\writefig	10.0	3.4	{\footnotesize $b(\rho)$}
\writefig	4.95	3.15	{\footnotesize $a$}
\writefig	3.1	1.75	{\footnotesize $c$}
\writefig	6.7	0.7	{\footnotesize $(-\mu-2dt)\rho$}
}
\caption{Upper and lower bounds for the ground state energy per site, $e_0(\rho)$. The
density that minimizes $e(\rho)$ necessarily satisfies $\rho\geq a$.}
\label{figbounddens}
\efig

We find an upper bound for $e_0(\rho)$ by using a variational argument. It is well-known that the
symmetric ground state is also the absolute ground state, so that we can consider a
non-symmetric trial function. We decompose $\Lambda$ into boxes of size $\ell =
\Lowerint{\rho^{-1/d}}$. We consider the trial function $\otimes_{j=1}^N \varphi_j$, where
$\varphi_j$ is supported in the $j$-th box only and minimizes the kinetic energy. As is
well-known, $\varphi_j$ is the ground state of the Dirichlet problem in the box, and the
corresponding eigenvalue is $-2dt \cos \frac\pi{\ell+1}$. Since $\ell+1 \geq \rho^{-1/d}$ and
$\cos x \geq 1 - \frac{x^2}2$, this eigenvalue is less than $-2dt + dt (\pi \rho^{1/d})^2$.
This implies that
\be
e_0(\rho) \leq b(\rho) \equiv (-\mu-2dt) \rho + \pi^2 dt \rho^{1+\frac2d}.
\ee
The minimum of $b(\rho)$ is reached for $\rho^{2/d} = \frac{\mu+2dt}{\pi^2 t (d+2)}$. The
minimum value is
\be
c = -\frac2{(\pi^2 t)^{\frac d2}} \Bigl( \frac{\mu+2dt}{d+2} \Bigr)^{1+\frac d2}.
\ee
By inspecting Fig.\ \ref{figbounddens} we find that the ground state density is
necessarily larger than
\be
a = \frac2{d+2} \Bigl( \frac{\mu+2dt}{\pi^2 t (d+2)} \Bigr)^{d/2}.
\ee
\end{proof}

\begin{proof}[Proof of Theorem \ref{thmnoMott}, (b)]
The strategy is the same as for part (a), although quantum fluctuations bring extra
complications. The variational argument leading to the upper bound for $e_0(\rho)$ can be
modified by replacing particles with holes, so as to yield
\be
e_0(\rho) \leq \tilde b(\rho) \equiv -\mu + (\mu-2dt) (1-\rho) + \pi^2 dt (1-\rho)^{1+\frac2d}.
\ee
The lower bound is trickier. We fix the density $\rho$ and work in the Hilbert space
$\caH_{\Lambda,N}$ with $N = \rho |\Lambda|$. We have that
\be
e_0(\rho) = -\lim_{\Lambda \nearrow \bbZ^d} \lim_{\beta\to\infty}
\tfrac1{\beta |\Lambda|} \log \Tr_{\caH_{\Lambda,N}} \e{-\beta(H-\mu N)}.
\ee
We can use the loop representation of Section \ref{secexpMott} for the trace to obtain an
expression similar to \eqref{partfctexpanded}; the difference is that we require the
sum of winding numbers of all loops to be equal to the negative of the number of holes $M =
|\Lambda|-N$.

The weights of loops with strictly positive winding numbers decays exponentially as
$\e{-\beta(U-\mu)}$, so they do not contribute in the limit $\beta\to\infty$. We obtain an
upper bound for $Z(\beta,\Lambda,\mu)$ (and therefore a lower bound for $e_0(\rho)$) by
neglecting the non-intersecting conditions between loops. Further, we replace the loops
$\gamma$ with negative winding numbers by trajectories $\theta$ as in the proof of Lemma
\ref{lemKPcrit} (c),(d). We then obtain the lower bound
\be
e_0(\rho) \geq -\mu\rho - \lim_{\Lambda \nearrow \bbZ^d} \lim_{\beta\to\infty}
\tfrac1{\beta |\Lambda|} \log \Tr_{\caH_{\Lambda,M}} \e{-\beta \tilde T} - \lim_{\Lambda
\nearrow \bbZ^d} \lim_{\beta\to\infty} \tfrac1{\beta|\Lambda|} \int_{\caL^{(0)}}
\dd\nu(\gamma) w(\gamma).
\ee
Here, $\tilde T$ denotes the multibody kinetic operator
\be
\tilde T = \sum_{x,y} \sigma(x-y) c_x^\dagger c_y,
\ee
and $\sigma(x)$ is given in \eqref{defsigma}. Then, by \eqref{uneautreborne},
\be
\lim_{\beta\to\infty} \tfrac1\beta \log \Tr_{\caH_{\Lambda,M}} \e{-\beta \tilde T} \leq
\bigl[ 2dt + 2^{10} d^2 \tfrac{t^2}U) \bigr] M.
\ee
The contribution of nonwinding loops is bounded using Lemma \ref{lemKPcrit} (a),
\be
\frac1{\beta|\Lambda|} \int_{\caL^{(0)}} \dd\nu(\gamma) w(\gamma) \leq \int_{\caL^{(0)}}
\dd\nu_0(\gamma) w(\gamma) \leq 2^{11} d \tfrac{t^2}U.
\ee
We have shown that
\be
e_0(\rho) \geq -\mu - (2dt - \mu + 2^{10} d^2 \tfrac{t^2}U) (1-\rho) - 2^{11} d
\tfrac{t^2}U.
\ee
From here on we proceed as before. The minimum of $\tilde b(\rho)$ is $-\mu - \frac2{(\pi^2 t)^{d/2}}
(\frac{2dt-\mu}{d+2})^{1+d/2}$. The ground state density then satisfies
\be
1-\rho \geq \frac{\frac2{(\pi^2 t)^{d/2}} (\frac{2dt-\mu}{d+2})^{1+d/2} - 2^{11} d
\frac{t^2}U}{2dt - \mu + 2^{10} d^2 \frac{t^2}U}.
\ee
One finds the condition of Theorem \ref{thmnoMott} by requiring that the numerator be
strictly positive.
\end{proof}

\appendix
\section{Cluster expansions}

This appendix contains the main theorem of \cite{Uel} for the convergence of cluster
expansions. It allows for an uncountable set of ``polymers'', so that it applies here.

Let $(\bbA, \caA, \mu)$ be a measure space with $\mu$ a complex measure. We suppose that 
$|\mu|(\bbA)<\infty$, where $|\mu|$ is the total variation (absolute value) of $\mu$.
Let $\zeta$ be a complex measurable symmetric function on $\bbA\times\bbA$.
Let $Z$ be the partition function:
\be
\label{deffpart}
Z = \sum_{n\geq0} \frac1{n!} \int\dd\mu(A_1) \dots \int\dd\mu(A_n) \prod_{1\leq
i<j\leq n} \bigl( 1 - \zeta(A_i,A_j) \bigr).
\ee
The term $n=0$ of the sum is understood to be 1.

We denote by $\caG_n$ the set of all (unoriented)
graphs with $n$ vertices, and $\caC_n \subset \caG_n$ the set of connected graphs of $n$ vertices.
We introduce the following combinatorial function on finite sequences $(A_1,\dots,A_n)$ of $\bbA$:
\be
\label{defphi}
\varphi(A_1,\dots,A_n) = \begin{cases} 1 & \text{if } n=1 \\ \frac1{n!} \sum_{G \in \caC_n} \prod_{(i,j)
\in G} [-\zeta(A_i,A_j)] & \text{if } n\geq2. \end{cases}
\ee
The product is over edges of $G$. A sequence
$(A_1,\dots,A_n)$ is a {\it cluster} if the graph with $n$ vertices and an edge between $i$ and $j$ whenever
$\zeta(A_i,A_j) \neq 0$, is connected.

Convergence of cluster expansion is garanteed provided the terms in \eqref{deffpart} are
small in a suitable sense. First, we assume that
\be
\label{condzeta}
|1-\zeta(A,A')| \leq 1
\ee
for all $A,A'\in\bbA$. Second, we need that the ``Koteck\'y-Preiss criterion'' holds true. Namely,
we suppose that there exists a nonnegative function $a$ on $\bbA$ such that for all $A\in\bbA$,
\be
\label{KPcrit}
\int\dd|\mu|(A') \, |\zeta(A,A')| \, \e{a(A')} \leq a(A),
\ee

The cluster expansion allows to express the logarithm of the partition function as a sum (or an
integral) over clusters.

\begin{theorem}[Cluster expansion]\hfill
\label{thmclexp}

\noindent
Assume that $\int\dd|\mu|(A) \e{a(A)} < \infty$, and that \eqref{condzeta} and
\eqref{KPcrit} hold true. Then we have
$$
Z = \exp\Bigl\{ \sum_{n\geq1} \int\dd\mu(A_1) \dots \int\dd\mu(A_n) \, \varphi(A_1,\dots,A_n) \Bigr\}.
$$
Combined sum and integrals converge absolutely. Furthermore, we have for all $A_1\in\bbA$
\be
\label{boundclusters}
1 + \sum_{n\geq2} n \int\dd|\mu|(A_2) \dots \int\dd|\mu|(A_n) \, |\varphi(A_1,\dots,A_n)| \leq
\e{a(A_1)}.
\ee
\end{theorem}

We refer to \cite{Uel} for the proof of this theorem, and for further statements about
correlation functions.

\end{document}